\pdfoutput=1
\documentclass[aip,jcp,reprint,citeautoscript]{revtex4-1}
\usepackage[utf8]{inputenc}
\usepackage[T1]{fontenc}

\usepackage[colorlinks=true,urlcolor=blue,linkcolor=blue,citecolor=blue]{hyperref}

\usepackage{amsfonts,amssymb,amsmath,amsbsy}
\usepackage{graphicx}
\usepackage{braket}
\usepackage[version=3]{mhchem}

\newcommand{\V}[1]{\boldsymbol #1}
\newcommand{\dee}{\text{d}}
\newcommand{\ie}{\emph{i.e.}}
\newcommand{\eg}{\emph{e.g.}}

\begin{document}

\title{Nanoplasmonics simulations at the basis set limit through %
completeness-optimized, local numerical basis sets}

\date{\today}

\author{Tuomas\ P.\ Rossi}
\email{tuomas.rossi@alumni.aalto.fi}
\affiliation{COMP Centre of Excellence, Department of Applied Physics, 
Aalto University School of Science, P.O.\ Box 11100, FI-00076 Aalto, 
Finland}

\author{Susi\ Lehtola}
\email{susi.lehtola@alumni.helsinki.fi}
\affiliation{COMP Centre of Excellence, Department of Applied Physics, 
Aalto University School of Science, P.O.\ Box 11100, FI-00076 Aalto, 
Finland}
\affiliation{Chemical Sciences Division, Lawrence Berkeley National 
Laboratory, Berkeley, California 94720, USA}

\author{Arto\ Sakko}
\affiliation{COMP Centre of Excellence, Department of Applied Physics, 
Aalto University School of Science, P.O.\ Box 11100, FI-00076 Aalto, 
Finland}
\author{Martti\ J.\ Puska}
\affiliation{COMP Centre of Excellence, Department of Applied Physics, 
Aalto University School of Science, P.O.\ Box 11100, FI-00076 Aalto, 
Finland}

\author{Risto\ M.\ Nieminen}
\affiliation{COMP Centre of Excellence, Department of Applied Physics, 
Aalto University School of Science, P.O.\ Box 11100, FI-00076 Aalto, 
Finland}
\affiliation{Dean’s Office, Aalto University School of Science, 
P.O.\ Box 11000, FI-00076 Aalto, Finland}

\begin{abstract}

We present an approach for generating local numerical basis sets of
improving accuracy for first-principles nanoplasmonics simulations
within time-dependent density functional theory. The method is
demonstrated for copper, silver, and gold nanoparticles that are of
experimental interest but computationally demanding due to the
semi-core d-electrons that affect their plasmonic response. The basis
sets are constructed by augmenting numerical atomic orbital basis sets
by truncated Gaussian-type orbitals generated by the
completeness-optimization scheme, which is applied to the
photoabsorption spectra of homoatomic metal atom dimers. We obtain basis
sets of improving accuracy up to the complete basis set limit and
demonstrate that the performance of the basis sets transfers to
simulations of larger nanoparticles and nanoalloys as well as to
calculations with various exchange-correlation functionals. This work
promotes the use of the local basis set approach of controllable
accuracy in first-principles nanoplasmonics simulations and beyond.

\end{abstract}

\maketitle

\section{Introduction}

Plasmonics attracts increasing interest due to its technological
relevance in numerous applications, such as biochemical sensing
\cite{Anker2008}, sub-wavelength light manipulation
\cite{Schuller2010}, and photovoltaics \cite{Atwater2010}. Plasmon
resonances in metal nano\-particles can be qualitatively understood by
classical electromagnetism, but the accurate description of
nanometer-size particles or systems with features in the subnanometer
range requires more elaborate approaches. In these systems the
plasmonic response is affected by quantum effects, such as electron
spill-out at the surface and electron tunneling. \cite{Halas2011} A
number of recent studies has demonstrated that the regime where these
effects start to play a role is experimentally
accessible. \cite{Savage2012,Scholl2012,Haberland2013,Scholl2013,Tan2014}

To study quantum effects in the plasmonic response computationally,
one often resorts to time-dependent density functional theory
\cite{Runge1984} (TDDFT) simulations. Qualitative understanding on the
quantum effects in nanostructures can be obtained within the jellium
approximation \cite{Ekardt1984,Puska1985,Zuloaga2009,Zuloaga2010}, but
it cannot capture the important atomic structure effects
\cite{Zhang2014}. In addition, the jellium model only describes simple
metals such as sodium, where the optical response is determined by the
valence s-electrons. However, the experimentally relevant materials
for plasmonics are usually coinage metals; copper (Cu), silver (Ag),
and gold (Au). 
\cite{Savage2012,Scholl2012,Haberland2013,Scholl2013,Tan2014,Kumara2011,Dharmaratne2014}
In these metals, in addition to the outermost s-electrons, also the
semi-core d-electrons participate in the response. Although the
effects due to the d-electrons can be accounted for in an approximate
manner in the jellium model \cite{Serra1997}, first-principles models
are necessary to obtain accurate results.

Coinage metal systems have been studied through numerous TDDFT
simulations, including metal nanoparticles of different shapes
\cite{Aikens2008,Stener2007,Guidez2013,Piccini2013,Bae2012a,Barcaro2014,%
Kuisma2014,Lopez-Lozano2014},
nanoalloys
\cite{Weissker2011,LopezLozano2013,Barcaro2011,Guidez2012,Liao2010},
protected metal clusters
\cite{Guidez2012a,Weissker2014,Malola2013,Malola2014,Gell2014}, and
nanoparticle dimers \cite{Bae2012}. However, TDDFT simulations for
these systems are computationally demanding. Even though the
calculations can be speeded up with the frozen-core approximation,
coinage metals require explicit calculation of the semi-core
d-electrons in addition to the s-electrons, resulting in 11 electrons
per atom in calculations, in contrast to, \eg, sodium where it usually
suffices to treat only the single 3s-electron per atom. Consequently,
simulated systems have typically been restricted to the maximum size
range of 100--200 coinage metal atoms
\cite{Aikens2008,Stener2007,Guidez2013,Piccini2013,Bae2012a,Weissker2011,%
LopezLozano2013,Barcaro2011,Guidez2012,Liao2010,Weissker2014,Gell2014,%
Malola2014,Guidez2012a,Bae2012},
with a few studies presenting larger systems, such as a \ce{Au263}
nanorod \cite{Lopez-Lozano2014}, a \ce{Ag272} nanoshell
\cite{Barcaro2014}, and a thiolate-protected \ce{Au314}
\cite{Malola2013}. The calculations have employed either real-space
grid codes
\cite{LopezLozano2013,Lopez-Lozano2014,Weissker2011,Weissker2014,%
Malola2013,Gell2014,Malola2014}
or the linear combination of atomic orbitals (LCAO) approach
\cite{Aikens2008,Stener2007,Guidez2013,Piccini2013,Bae2012a,Barcaro2014,%
Barcaro2011,Guidez2012,Liao2010,Guidez2012a,Bae2012}. Recently,
a new LCAO-TDDFT implementation was developed \cite{Kuisma2014},
allowing to push the accessible system size close to the classical
limit (\ce{Ag561} presented in Ref.~\onlinecite{Kuisma2014}).

A serious problem of the LCAO approach is that it is prone to errors
due to basis set incompleteness --- a problem which can be
straightforwardly tackled in the real-space grid
methods. Nevertheless, this issue has not been extensively discussed
in previous nanoplasmonics studies using the LCAO approach. Instead, a
reasonable accuracy of the results has been checked by calculations of
test systems with larger basis sets of the available basis set series
\cite{Stener2007,Aikens2008,Barcaro2014} or by comparing to real-space
grid results \cite{Kuisma2014}.

The basis set issue is complicated by the fact that conventional basis
sets are typically optimized for ground-state energy
calculations. \cite{Jensen2013} For other properties, such as dipole
moments, excited states, or plasmonic charge density oscillations,
these basis set series are not expected to yield quickly converging
results
\cite{Miura2007,Chong1995,Manninen2006,Lehtola2012Completeness,Lehtola2015}.
In the case of photoabsorption spectra, the accurate description of
the dipole moment of the excited states is essential. This
necessitates inclusion of diffuse functions (\ie, functions with large
spatial extent) in the basis set. Diffuse functions are not present in
the energy-optimized basis sets because of their minor contribution to the
ground-state energy of electrically neutral systems, but they can be
generated into these basis sets by, \eg, minimizing the energy of
anions. \cite{Jensen2013} Alternatively, unoccupied atomic orbitals
can be included in the basis set, which has been found to improve 
the description of the photoabsorption of metal 
nanoparticles 
\cite{Kuisma2014,Tsolakidis2002}. 
However, these
approaches are not guaranteed to be optimal for extending the basis
sets beyond conventional ground-state energy calculations.

In the present work, we show that efficient basis sets specifically
optimized for describing the plasmonic optical response can be
systematically generated using the completeness-optimization (CO)
approach \cite{Manninen2006}. CO is a black-box procedure for
generating basis sets for any property at any level of theory. It has
been previously used to generate all-electron Gaussian-type orbital
(GTO) basis sets for calculating magnetic
\cite{Manninen2006,Ikalainen2009,Lantto2011,Vahakangas2013,Abuzaid2013,%
Jaszunski2013,Vaara2013}
and magneto-optic
\cite{Ikalainen2008,Ikalainen2010,Ikalainen2012,Pennanen2012,Shi2013,%
Fu2013,Vahakangas2014}
properties as well as the electron momentum density
\cite{Lehtola2012Completeness,Lehtola2013Contraction}. In this work,
we present a straightforward extension of the CO formalism to
semi-numerical basis sets by combining numerical atomic orbitals
(NAOs) with truncated numerical Gaussian-type orbitals (NGTOs). The
NGTOs are selected by a recently developed automatic CO procedure
\cite{Lehtola2012Completeness,Lehtola2015} to augment NAO basis sets
with the necessary diffuse and polarization functions. We demonstrate
the applicability of the scheme for describing collective plasmonic
excitations in coinage metal clusters.  We optimize the basis sets to
reproduce the photoabsorption spectra of homoatomic dimers, and show
that the generated basis sets are transferable to larger nanoparticles
and to different chemical environments in nanoalloys, as well as to
different exchange-correlation functionals.

The paper is organized as follows. In Sec.\ \ref{sec methods}, we give
an overview of the used methodologies --- TDDFT, LCAO, and CO. In
Sec.\ \ref{sec implementation}, we describe our implementation, and in
Sec.\ \ref{sec results} we demonstrate the performance of the basis
set generation and test the transferability of the generated basis
sets. We conclude the study in Sec.\ \ref{sec conclusions}.

\section{Methods}
\label{sec methods}

\vspace{-2mm}

\subsection{Time-dependent density functional theory}
\label{sec tddft}

\vspace{-1mm}

TDDFT is a well-established formulation of the time-dependent
many-body Schrödinger equation in terms of the time-dependent electron
density. \cite{Runge1984} The theory is usually applied within the
Kohn--Sham (KS) description of density functional theory (DFT)
\cite{Hohenberg1964,Kohn1965}, which models the interacting
many-electron system as a non-interacting system in an effective
potential. In this approach, the complicated many-body interactions
are described by the so-called exchange-correlation (xc) functional.
The time-dependent xc functionals are usually treated in the adiabatic
limit, \ie, an instantaneous time-dependent density is used as input
for the ground-state functional. \cite{Marques2012}

The dynamical response, and in particular, the photoabsorption
spectrum of a given system, can be calculated in two formally
different, but equivalent manners within the TDDFT
framework. First, it can be obtained from the time-dependent dipole
moment that is recorded during the explicit real-time propagation of
the KS-orbitals that have been excited from the ground state by a
$\delta$-pulse perturbation \cite{Yabana1996}. Second, the excitations
of the system can be calculated by formulating the linear density
response to an external perturbation in the frequency space, yielding
the Casida matrix equation \cite{Casida1995}.

In this work, we use both the time-propagation and Casida schemes for
calculating photoabsorption spectra. We employ the open source GPAW
program
\cite{Mortensen2005,Walter2008,Larsen2009,GPAW,Enkovaara2010,Bahn2002}
in TDDFT calculations. GPAW uses the projector augmented wave (PAW)
method \cite{Blochl1994} for freezing the inert core electrons and for
obtaining smooth pseudo-wave functions in the vicinity of the
nuclei. The simulations explicitly include only the outermost
electrons, \ie, for the coinage metals, the semi-core d-electrons and
the valence s-electrons (11 electrons per atom in total). 
The element-specific PAW transformations are constructed at the 
scalar-relativistic level of theory. Thus, 
relativistic effects, especially important for gold \cite{Pyykko2012},
are included implicitly in the calculations 
through the PAW transformation.
For the
present study, GPAW has the advantage that it can describe wave
functions either on a real-space grid \cite{Mortensen2005,Walter2008}
or within the LCAO approach \cite{Larsen2009,Kuisma2014}. In both
modes, uniform real-space grids are used for representing electron
densities and potentials. These two modes of operation share a
significant portion of computational framework within the program,
which allows us to compute grid-based spectra and LCAO spectra with
minimal sources of differences apart from the representation of the
wave functions.

\subsection{Linear combinations of atomic orbitals}
\label{sec lcao}

In the LCAO approach the single-electron KS wave function is expressed
as
\begin{align}
\label{eq lcao}
\ket{\psi} = \sum_{a}^{\text{atoms}} \sum_{\nu=1}^{N_a} c^a_{\nu}
\ket{\chi^a_{\nu}} ,
\end{align}
where $c^a_{\nu}$ are the expansion coefficients for basis functions
$\ket{\chi^a_{\nu}}$ centered on the atom $a$, $N_a$ denoting the
amount of basis functions on that atom. In a coordinate
system centered on the atom $a$, an associated basis function is
written as a product of a radial function $\phi^a_{nl}(r)$ and a
spherical harmonic $Y_{lm}(\theta, \varphi)$,
\begin{align}
\label{eq bf}
\braket{\V r | \chi^a_{\nu}} = \chi^a_{\nu}(\V r) =
\phi^a_{n_{\nu}l_{\nu}}(r) Y_{l_{\nu}m_{\nu}}(\theta, \varphi) .
\end{align}
Above, $\nu$ is a symbolic index over the combinations of $n_{\nu}$, $l_{\nu}$,
and $m_{\nu}$. In this work, radial functions are taken to be either NAOs or
NGTOs as described in Sec.~\ref{sec implementation}.

The main advantage of the LCAO approach is that a sufficiently
accurate description of the wave function can often be achieved with a
small number of basis functions. In addition, in the case of truncated
basis functions, the number of overlapping basis functions is usually
small, which enables efficient computations. The main drawback of the
LCAO approach is that it is prone to errors due to the incompleteness
of the basis set. Thus, it is important to use basis sets that are
flexible enough for describing wave functions accurately in the
regions that are essential for the property in question.

\vspace{-2mm}

\subsection{Completeness-optimization}
\label{sec co-opt}

\vspace{-2mm}

CO is a general approach for generating optimal basis sets for any
chosen property. \cite{Manninen2006} The method is based on the
concept of the completeness profile \cite{Chong1995} that is defined
as
\begin{align}
\label{eq profile}
Y_l(\alpha) = \sum_{\mu,\nu = 1}^{N} \braket{g_l(\alpha)|\chi_{\mu}}
S_{\mu\nu}^{-1} \braket{\chi_{\nu}|g_l(\alpha)} ,
\end{align}
where $\ket{\chi_{\nu}}$ are the basis functions, $S_{\mu\nu}^{-1}$ is
the $(\mu, \nu)$ element of the inverse overlap matrix
orthonormalizing the basis, and $g_l(\alpha)$ is a primitive test
function usually taken to be a normalized Gaussian primitive
$g_l(\alpha) \propto r^l e^{-\alpha r^2} Y_{lm}(\theta,\varphi)$. For
basis functions of the form of Eq.~\eqref{eq bf}, the inner product
$\braket{g_l(\alpha)|\chi_{\nu}} \propto \braket{r^{l_{\alpha}}
  e^{-\alpha r^2} | \phi_{n_{\nu}l_{\nu}}} \delta_{l_{\alpha}l_{\nu}}
\delta_{m_{\alpha}m_{\nu}}$.  Thus, each value of the angular momentum
$l$ yields a different completeness profile $Y_l(\alpha)$, whereas all
$m$ values of a given $l$ yield the same profile.

The profile essentially measures the validity of the resolution of the
identity operator
\begin{align}
\sum_{\mu,\nu = 1}^{N} \ket{\chi_{\mu}} S_{\mu\nu}^{-1}
\bra{\chi_{\nu}} \approx \mathbb{I} ,
\end{align}
which would be exact for a complete basis set. Correspondingly, the
completeness profile satisfies \mbox{$0 \leq Y_l(\alpha) \leq 1$}.

The idea in CO is to generate a basis set that has $Y_l(\alpha)
\approx 1$ within the intervals $[\alpha_{l}^{\text{min}},
  \alpha_{l}^{\text{max}}]$ that are important for the property in
question. 
This is accomplished by optimizing the parameters of the basis functions 
so that $Y_l$ is maximized within the intervals. The number of basis 
functions needed for this depends on the tolerance $\tau_l$ for
deviations from unity in $Y_l$ within the intervals.
\cite{Manninen2006,Lehtola2012Completeness,Lehtola2015} 
The task of a practical CO algorithm is to find the optimal limits of 
the intervals, while keeping $Y_l(\alpha) \approx 1$ within the intervals.

In this work, we use Gaussian basis
functions characterized by their exponents $\alpha$ 
and employ the automatic CO procedure
\cite{Lehtola2012Completeness,Lehtola2015} implemented in the open
source ERKALE program \cite{Lehtola2012ERKALE,ERKALE}. The procedure
is based on systematic trial and error searches for determining the
complete basis set (CBS) limit of the property, as well as the CBS
itself. The algorithm is divided in two phases. First, in the
extension phase, the CBS is found by progressively extending the basis
set with the \emph{most important} functions until the property is
converged. Each addition of a basis function corresponds to an extension 
of the intervals $[\alpha_{l}^{\text{min}}, \alpha_{l}^{\text{max}}]$ or 
decrease of the tolerance $\tau_l$.
Second, in the reduction phase, the \emph{least important}
basis functions are repeatedly pruned from the basis, shrinking the 
$\alpha_l$ intervals or increasing $\tau_l$.
During the optimization, the relative importance of 
the addition or removal of a basis
function is determined by a user-defined error metric.
In every step of the algorithm, the exponents of the Gaussian basis functions
are determined by maximizing the completeness profile $Y_l$ within the
current $\alpha_l$ interval.
As a result of the reduction phase, a
systematic sequence of basis sets of decreasing size and accuracy is obtained. 

\section{Implementation}
\label{sec implementation}

\vspace{-3mm}

\subsection{Application of the completeness-optimization in nanoplasmonics}
\label{sec application}

\vspace{-2mm}

The CO routine \cite{Lehtola2015} has a general interface for basis
set generation. We have written a wrapper program that implements this
interface and calls the GPAW program \cite{Enkovaara2010} for
photoabsorption spectrum calculations. The practical workflow between
the programs is as follows. The optimization routine forms trial GTO
basis sets and feeds them to the wrapper. The wrapper transforms the
GTO sets into NGTOs, prepares the input for GPAW, and submits the GPAW
calculations. Once the GPAW jobs have completed, the wrapper reads in
the results and returns them to the optimization routine, which
interprets them through the error metric and uses the information to
update the basis set and generate new trials.

The error metric for determining the effect of modifying the basis set
is defined as follows. The error $\epsilon$ of a photoabsorption
cross-section $\sigma_{\text{abs}}(\omega)$ from the corresponding
reference spectrum $\sigma_{\text{abs}}^{\text{ref}}(\omega)$ (given
by an earlier CO step) is defined as
\begin{align}
\label{eq error def}
\epsilon = \min_\delta \left[ \frac{\displaystyle
    \left(\int_{\omega_{\text{min}}}^{\omega_{\text{max}}}
    |\sigma_{\text{abs}}(\omega-\delta) -
    \sigma_{\text{abs}}^{\text{ref}}(\omega)|^2 \dee \omega
    \right)^{1/2} }{\displaystyle
    \left(\int_{\omega_{\text{min}}}^{\omega_{\text{max}}}
    |\sigma_{\text{abs}}^{\text{ref}}(\omega)|^2 \dee \omega
    \right)^{1/2}} \right. \nonumber \\
+ e^{(\delta/\gamma)^2} -1\Bigg] ,
\end{align}
where $\omega_{\text{min}}$ and $\omega_{\text{max}}$ are cut-off
energies, $(e^{(\delta/\gamma)^2}-1)$ is a penalty term for a constant
energy shift $\delta$ in the spectrum, and $\gamma$ describes the
stiffness of the penalty function. The error metric also depends
implicitly on the parameters used to broaden the discrete TDDFT
spectrum to model the finite lifetime of excitations as well as
temperature effects and instrumental resolution. 
Without explicitly allowing an energy shift $\delta$ 
in the definition, the error measure 
would be much more sensitive to offsets in energy than to
changes in intensity. The energy shift $\delta$ is penalized
through the penalty term, in which the parameter $\gamma$
 provides a sliding scale to 
balance the sensitivity of the measure between energy and intensity.
With $\gamma \to 0^+$ the penalty term becomes extremely stiff and
$\delta = 0$ always minimizes the error metric, 
in which case the metric becomes
the usual normalized L$^2$ measure.
The error metric is also used for benchmarking
the quality of the obtained basis sets. In this case, the reference
spectrum is obtained from a real-space grid calculation. However, it
is emphasized that during the optimization the grid reference is not
employed by any means.

A practical issue concerning the extension of the CO routine to the
frozen-core approximation is that the CO routine was initially designed for
all-electron calculations. In the present work, the basis sets only represent 
the outermost s- and d-electrons of the coinage metal atoms, 
as the core electrons are described implicitly.
Thus, the assumptions made for the composition of the basis
set in the all-electron case \cite{Lehtola2015} have been relaxed
here, allowing for a non-monotonic amount of GTOs on consecutive
angular momentum shells during basis set reduction.

\subsection{NAO+NGTO basis sets}
\label{sec nao+ngto}

The benefit of NAOs is that they are not restricted to be of any
analytical form. Thus, they should excel in the highly structured
atomic core region, which is only slightly affected by the surrounding
chemical environment. Although the PAW transformation results in a
smoother pseudo-wave function in this region, the NAOs are still
pre-eminent for describing the atomic ground state.

While the generation of minimal-basis NAOs is straightforward, the
case for polarization and multiple-$\zeta$ functions
 is not as clear. NAOs do not hold
advantages for these functions, as the forms of the 
functions are not generally known and the polarization changes from
one system to the other. NAO polarization and multiple-$\zeta$
 functions are typically
generated from gas-phase atomic wave functions by splicing the radial
function \cite{Soler2002} and/or by including bound unoccupied orbitals, 
but also other
systematic approaches for generating NAO basis sets have been developed 
\cite{Blum2009,Ren2012,Zhang2013,Corsetti2013}.
Here, (N)GTOs
have a definite advantage, as systematic sets of functions of any
angular momentum can easily be generated. 
In the default basis sets of GPAW, NGTOs are employed in addition 
to numerical orbitals and splicing \cite{Larsen2009}.

We employ the following strategy for basis set generation. NAOs are
used to represent the atomic ground state, for which they hold the
definitive advantage. This minimal basis is then augmented with NGTOs
that are generated by the CO routine, adding the desired polarization
and diffuse functions.

The CO routine treats analytical GTOs, whereas in this work, numerical
basis functions are used. Thus, the analytical GTOs are smoothly 
truncated to NGTOs in the wrapper program by a second-order 
polynomial so that the radial part becomes
\begin{align}
G(r) = A r^l \big( e^{-\alpha r^2} - (a - b r^2)\big) ,
\end{align}
where $A$ is a normalization constant, and the constants $a$ and $b$
are chosen so that $G(r_{\text{c}}) = G'(r_{\text{c}}) = 0$ at the
cut-off radius $r_{\text{c}}$. \cite{NoteNGTO} The cut-off radius is
determined by requiring that the NGTO differs point-wise from the
exact GTO $\propto r^l e^{-\alpha r^2}$ at most by $10^{-3}$
au$^{-3/2}$. This strict criterion distorts the Gaussian shape
negligibly, but, on the other hand, leads to large $r_{\text{c}}$
values, which has a detrimental effect on the computational cost.

Because of the negligible difference between GTOs and NGTOs, the
analytical form is used in the optimization of the
Gaussian primitives, \ie, in the maximization of the completeness
profile Eq.~\eqref{eq profile} determining the exponents of the 
primitives. 
Additionally, the underlying NAOs are
neglected in Eq.~\eqref{eq profile}. Although NAOs (and truncated NGTO
forms) could be straightforwardly included in the optimization of the
completeness profile (Eq.~\eqref{eq profile}), the effects due to the
explicit account of the NAOs at this stage are expected to be small
due to the different asymptotic forms of NAOs and GTOs. The
incorporation of the underlying NAO basis set is done in the wrapper
program so that its presence is invisible to the CO routine and the
major effect of NAOs rises implicitly through the spectrum
calculation.

The following \emph{ad hoc} restrictions are imposed for the allowed
NGTOs. At least 90\% of the norm of the function must be within a
sphere of radius 6~\AA\ around the nucleus and at most 90\% of the
norm may be within 0.6~\AA. \cite{NoteRestr} The first condition
results in the rejection of extremely diffuse functions that would
require impractically large simulation grids, and which would cause
severe numerical problems due to linear dependencies in extended
systems.  The second condition ensures that even the tightest
functions can be faithfully mapped to the real-space grid used to
describe their contributions to the electron density. However, tight
functions are not usually needed anyhow because of the PAW transformation.

\subsection{Numerical parameters}
\label{sec parameters}

The CO was started from an initial NAO+NGTO basis set composed of
three radial functions on the s-, p-, and d-shells, totaling 27 basis
functions per atom. The initial GTO basis set was energy-optimized
\cite{Lehtola2015} for the gas-phase atom in question.
The value $\gamma = 0.4$~eV was used in the error
metric (Eq.~\eqref{eq error def}) during the CO, and the spectra were
broadened with Gaussians using a full width at half maximum (FWHM) of
$0.47$~eV. 
The FWHM and $\gamma$ parameters used in the CO were determined by
trial and error to obtain a suitable balance between the energy 
and intensity sensitivities. The parameters were not specifically
optimized and other values that achieve a proper balance
are expected to yield similar results.
For the integration limits, $\omega_{\text{min}} = 0$~eV
was used, and $\omega_{\text{max}}$ was set to 5.2~eV, 5.4~eV, and
6.7~eV for \ce{Cu2}, \ce{Ag2}, and \ce{Au2}, respectively. The
$\omega_{\text{max}}$ limits were estimated from the grid-reference
calculations so that spurious box-state transitions are excluded. The
same parameters are also used in Sec.~\ref{sec generation} for
calculating the errors with respect to the grid reference.
The extension phase was terminated when the
error between consecutive spectra was smaller than $0.011$. The
maximum angular momentum in the basis set was set to $l=3$, allowing
for s-, p-, d-, and f-type GTOs to be generated by the
algorithm. The effect of higher-$l$ basis functions is expected to
be insignificant.

The LCAO spectra of homoatomic dimers were calculated within Casida's
linear-response TDDFT formalism \cite{Casida1995} by including the
full eigenstate spectrum corresponding to the finite basis set and
averaging over the longitudinal and transversal components. Other
systems and all grid references were calculated with time-propagation
TDDFT \cite{Yabana1996} using a weak $\delta$-pulse and a time step of
$10$~as. The Perdew-Burke-Ernzerhof (PBE) xc functional \cite{Perdew1996,Perdew1997} in the
adiabatic limit was used in all the calculations, unless otherwise
stated.  All the calculations were done as spin-paired. The following
GPAW-specific parameters were used. LCAO mode: grid spacing
$h=0.3$~\AA\ and minimal vacuum size around the system $d_{\text{vac}}
= 6$~\AA. Grid mode: $h=0.25$~\AA\ and $d_{\text{vac}} = 8$~\AA.
The accuracy of the Hartree potential evaluation within the simulation cell was
improved by employing multipole corrections to the potential. \cite{Castro2003}
The
grid mode uses the real-space grid for describing the wave functions,
which explains the smaller $h$ and larger $d_{\text{vac}}$ values
needed for converged results. In the LCAO mode, the numerical basis
functions are described on their specific radial grids and the uniform
real-space grid is used only for representing the density and
potentials. \cite{Larsen2009}

In the transferability tests (see Secs.~\ref{sec transf}--\ref{sec xc}),
the spectra were convoluted by a
Gaussian broadening with FWHM = $0.20$~eV.
The value $\gamma = 0.25$~eV was used in the error metric to obtain
a reasonable energy/intensity sensitivity.
These parameters yield the same error for a spectrum shift
of $0.1$~eV and an intensity change by 20\% for a test spectrum with a
single absorption peak. The integration limits in the error metric
were set to $\omega_{\text{min}} = 0$~eV and $\omega_{\text{max}} =
5$~eV to probe the visible--near-UV region of spectrum.

\vfill

\section{Results}
\label{sec results}

\subsection{Generation of basis sets}
\label{sec generation}

The basis sets were generated independently for Cu, Ag, and Au by
using the photoabsorption spectrum of the corresponding homoatomic metal atom
dimer as the completeness-optimization target. The dimer bond lengths
were optimized with GPAW by using the default dzp basis sets and the
PBE functional. The obtained bond lengths for \ce{Cu2}, \ce{Ag2}, and
\ce{Au2} are 2.23~\AA, 2.58~\AA, and 2.55~\AA, respectively. The used
values agree well with the experimental values, 2.22~\AA, 2.53~\AA,
and 2.47~\AA, respectively \cite{Lombardi2002}.

\begin{figure}[b]
\centering \includegraphics{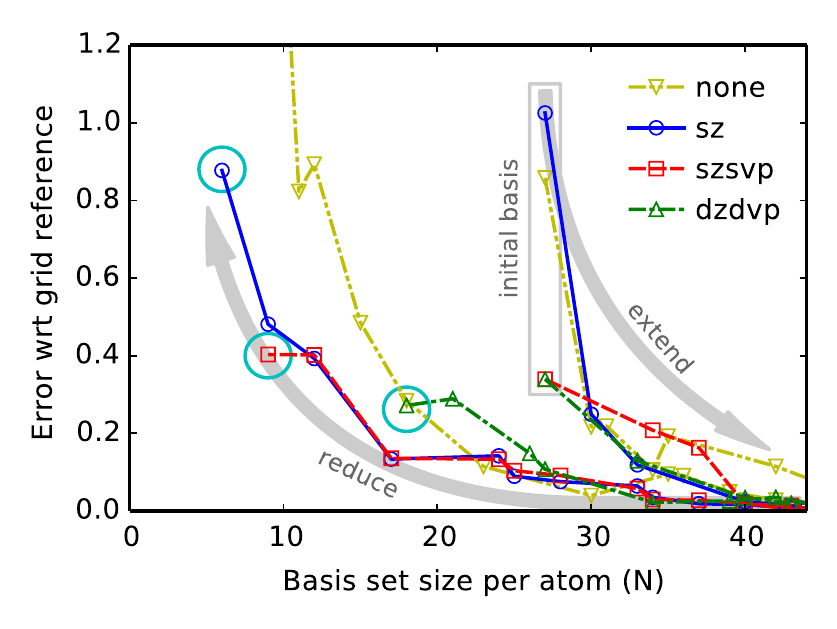}
\caption{The progression of the CO procedure for \ce{Ag2} during the
  extension and reduction phases as illustrated by the error with
  respect to the grid-calculated reference spectrum (Eq.\ \eqref{eq
    error def}). Different underlying NAO basis sets are used: 1)
  ``none'': only NGTOs, 2) ``sz'': single-$\zeta$ basis of 4d and 5s
  orbitals, 3) ``szsvp'': single-$\zeta$ basis of 4d, 5s, and 5p
  orbitals, 4) ``dzdvp'': double-$\zeta$ basis of 4d, 5s, and 5p
  orbitals. The cyan circles mark the NAO-only basis sets.}
\label{fig dimer}
\end{figure}

The progression of the CO procedure for the silver dimer is
illustrated in Fig.~\ref{fig dimer} for different choices of the
underlying NAO basis sets.  In the extension phase, the error in the
LCAO calculation decreases rapidly. Once the complete basis set (CBS)
limit has been reached, the least-important primitives are pruned out
one by one in the reduction phase.
During many sequential steps, the reduction-phase basis
sets yield more accurate results with less basis functions than the
ones from the extension phase. The progression of the algorithm is not
completely monotonic, because the optimized property is not
variational. \cite{Lehtola2015}

For the rest of the work, we focus on the NAO-sz+NGTO basis
sets. \cite{NoteSuppl} Then, only a minimal NAO basis set is
included, so that CO produces all the polarization functions necessary
for describing the chemical ground-state environment as well as the
excited state characteristics. Contrary to NAO-only basis sets, the
NAO-sz+NGTO basis sets are completely general. The occupied orbitals
used in the NAO-sz basis can be generated for any element, whereas the
first unoccupied orbitals may not be bound, in which case they cannot be
directly included. Additionally, the comparison between ``sz'' and
``szsvp'' reduction series in Fig.~\ref{fig dimer} indicates that the
use of the unoccupied p-orbital in the NAO basis does not result in a
large difference, at least for the silver dimer.

Furthermore, the NAO-sz+NGTO basis sets are expected to be numerically
efficient. We note from Fig.~\ref{fig dimer} that these basis sets
indeed perform better than either the NAO-dzdvp+NGTO sets or the pure
NGTO sets. The reasons are that the NAO-dzdvp basis sets contain extra
NAO functions, which are not free to be optimized by the CO method,
and that the pure NGTO basis sets do not have the advantage of the
minimal NAO-sz basis sets. However, the NAO basis sets depend on the
used xc functional whereas the pure NGTO basis sets could be more
transferable across different functionals. Still, the underlying NAOs
can be easily changed with the functional. This approach is presented
in Sec.~\ref{sec xc}.

\begin{figure}[b]
\centering \includegraphics{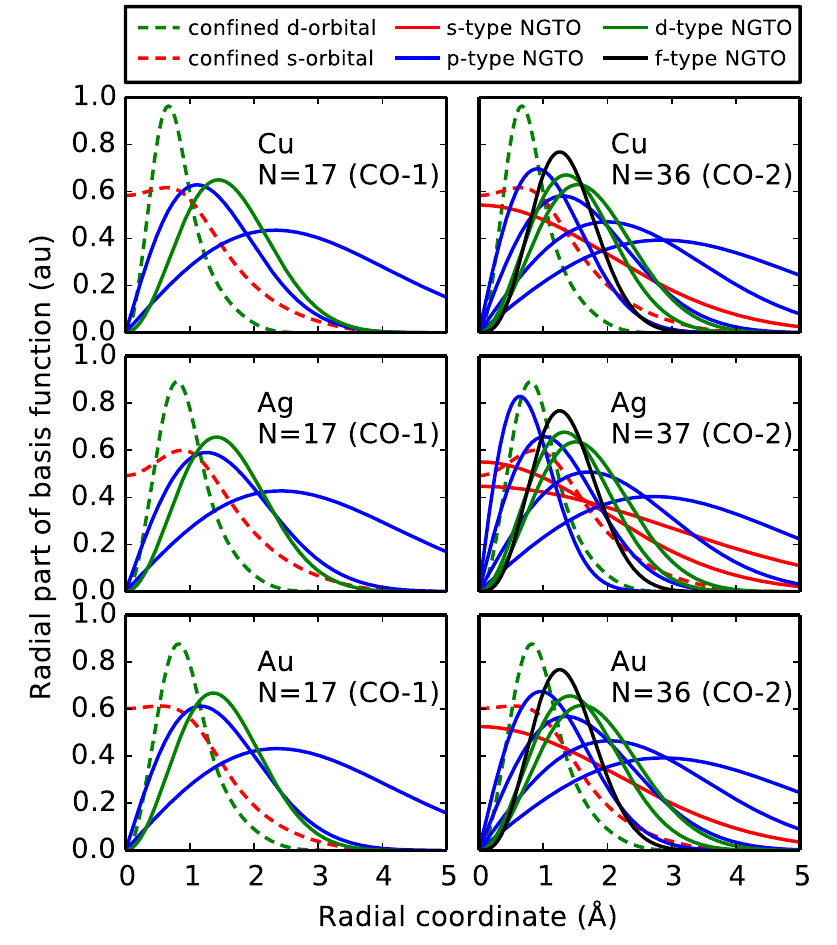}
\caption{NAO-sz+NGTO basis sets generated by the CO algorithm. Basis
  sets with 17 and 36 or 37 functions for Cu, Ag, and Au are shown.}
\label{fig basis}
\end{figure}

To illustrate the generated basis sets, we present two NAO-sz+NGTO
basis sets for Cu, Ag, and Au in Fig.~\ref{fig basis}. The basis sets
with $N=17$ are expected to yield decently accurate results (see
Fig.~\ref{fig dimer}) and the $N=36$ / $N=37$ ones are close to the
CBS limit. Henceforth, we refer to the $N=17$ basis sets as ``CO-1''
and the $N=36$ / $N=37$ basis sets as ``CO-2''. The similarity of the
basis sets across the studied elements is evident. In the CO-1 basis
sets, the NAO-sz basis is augmented by two diffuse p-type NGTOs and a
single d-type NGTO. In the CO-2 case, additional NGTOs are included on
all shells. The Gaussian exponents of the NGTOs are similar across all
studied elements, which is expected due to their closely related
chemical characteristics. Yet, even though the shown basis sets are
similar for all elements, the whole basis set \emph{series} are not
the same, as functions are pruned out in different orders in the reduction
phase. For example,
there is no $N=36$ basis set for silver. Note also that
the most diffuse p-type NGTOs are at the constraint limit imposed for NGTOs
(see Sec.\ \ref{sec nao+ngto}).

\subsection{Transferability of basis sets to larger clusters}
\label{sec transf}
\vspace{-2mm}

\begin{figure}[h!]
\centering \includegraphics{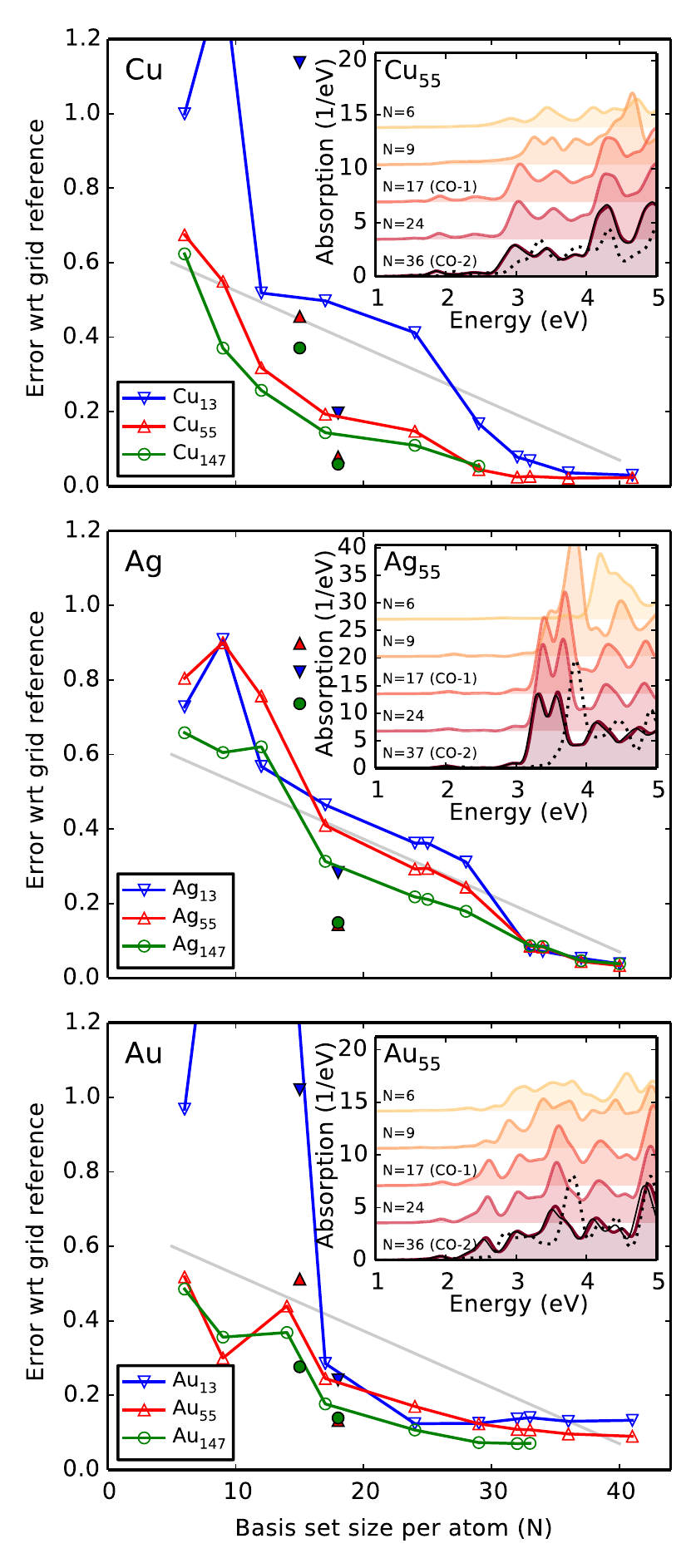}
\caption{Transferability of the generated NAO-sz+NGTO basis sets to
  icosahedral homoatomic coinage metal clusters of 13, 55, and 147
  atoms. The off-line filled markers indicate results calculated with
  the GPAW default dzp ($N=15$) and NAO-dzdvp basis sets ($N=18$). The
  straight gray line is drawn to ease the visual comparison. The
  insets present photoabsorption spectra of 55-atom clusters
  calculated with basis sets of increasing size. The grid reference
  (thin black line) and the LCAO spectrum with the default dzp basis
  (dotted line) are also shown in the insets.}
\label{fig transf}
\vspace*{-4ex}
\end{figure}

The usefulness of basis sets depends on their transferability to
different chemical environments. Here, we consider the transferability
of the basis sets obtained in the previous section to larger systems
by using homoatomic icosahedral clusters of 13, 55, and 147 atoms as
test cases. The clusters were constructed by adding icosahedral Mackay
layers one by one around a central atom. The structures were relaxed
with GPAW by using the default dzp basis sets and the PBE functional,
but their icosahedral symmetries were not significantly
disturbed. \cite{NoteSuppl}

We show in Fig.~\ref{fig transf} the error in the photoabsorption
spectra (Eq.~\eqref{eq error def}) for the clusters. We observe a
nearly monotonic increase in the accuracy with increasing basis set
size. The magnitude of the error is similar between different
elements, and the error tends to decrease when the system size
increases.

In Fig.~\ref{fig transf}, we also show for comparison the errors of the
GPAW default dzp basis set and the NAO-dzdvp basis set that has been
used in a previous study \cite{Kuisma2014}. The default dzp basis set
is unsuitable for describing the response, which is due to its lack of
diffuse p-functions. \cite{Kuisma2014} The NAO-dzdvp basis set
provides an equivalent or better accuracy than the NAO-sz+NGTO basis
set of similar size. However, in contrast to the NAO-only basis sets,
the basis sets generated in the present work allow for further,
systematic improvements in accuracy beyond that of the NAO-only basis
sets.

The insets in Fig.~\ref{fig transf} illustrate how the spectra look
for the 55-atom clusters calculated with basis sets of different
size. We observe that all the spectral features are mostly correct
with the basis sets of $17$ or more functions per atom. The CO-1
spectra suffer from a blue-shift of $0.1$--$0.2$~eV and the largest
improvement in the spectrum when growing the size of the basis set
comes from a red-shift towards the converged spectrum. The CO-2 basis
sets of Cu and Ag are already at the CBS limit, as the positions of the
spectral peaks coincide with the grid references within $0.05$~eV. For
Au, the convergence is slow after $N\approx24$, but also there the
CO-2 spectrum is near the grid reference except for a few details.

\subsection{Transferability of basis sets to nanoalloys}
\label{sec alloys}

Now, we consider the transferability of the basis sets that were
optimized for homoatomic dimers to heterogeneous metal clusters, where
the chemical environments are different from the homogeneous
systems. As test systems, we take the core-shell clusters
\ce{Ag13Cu42}, \ce{Cu13Ag42}, \ce{Ag13Au42}, and \ce{Au13Ag42}, which
consist of icosahedral 13-atom cores and single 42-atom Mackay layers
around the cores, as well as two icosahedral alloys, \ce{Cu14Ag20Au21}
and \ce{Cu18Ag17Au20}, which were generated by random substitution of
atoms in the 55-atom icosahedral geometry. The clusters were relaxed
analogously to the homoatomic clusters in Sec.~\ref{sec transf}.
\cite{NoteSuppl}

\begin{figure}[b]
\centering \includegraphics{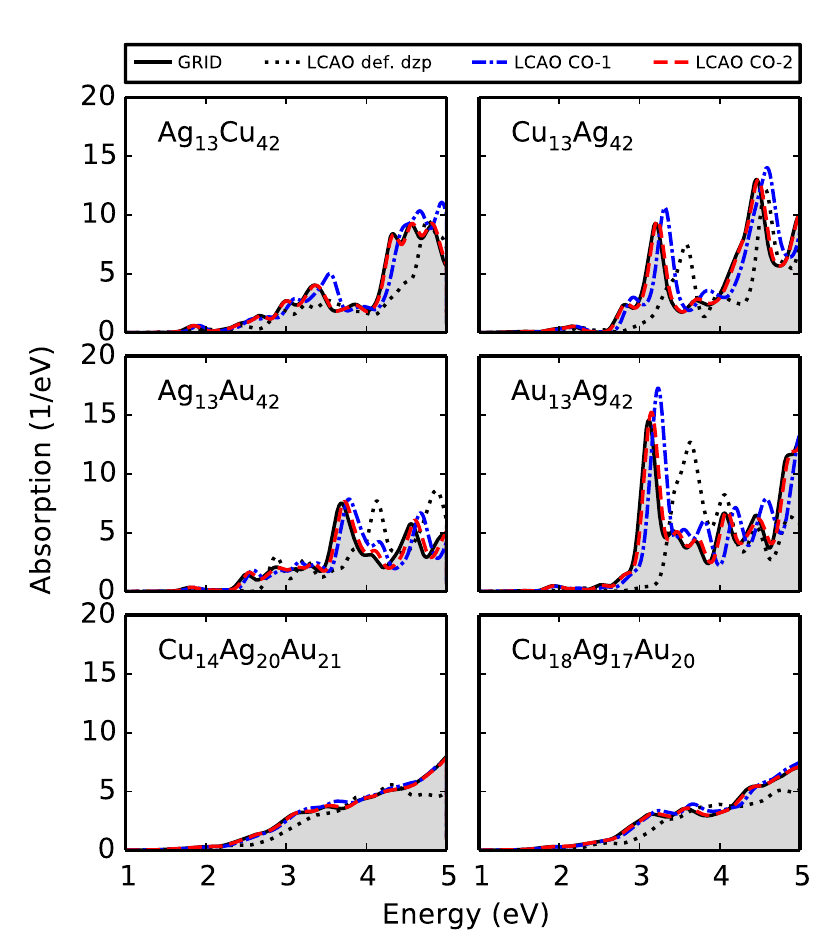}
\caption{Transferability of the basis sets to alloy clusters. The
  gray shading is applied to the grid-reference spectra.}
\label{fig alloys}
\end{figure}

The photoabsorption spectra of the alloy clusters are shown in
Fig.~\ref{fig alloys}. The CO-2 spectra are again in excellent
agreement with the grid reference. The smaller CO-1 basis set results
in a $0.1$--$0.2$~eV blue-shift of the spectra. Due to the lower
symmetry and breaking of degeneracies the disordered clusters have
fewer sharp spectral features than the core-shell clusters. This is
also reflected in that only small differences between the CO-1 and
CO-2 spectra can be seen.

The spectra calculated with the default dzp basis sets are also shown
in Fig.~\ref{fig alloys}. As in the case of homoatomic clusters, these
basis sets are unable to sufficiently describe the photoabsorption
spectra. The trends in the results obtained with the NAO-dzdvp basis
sets (not shown) are similar to those in the homoatomic case.

\subsection{Transferability of basis sets to different xc functionals}
\label{sec xc}

\begin{figure}[h]
\centering \includegraphics{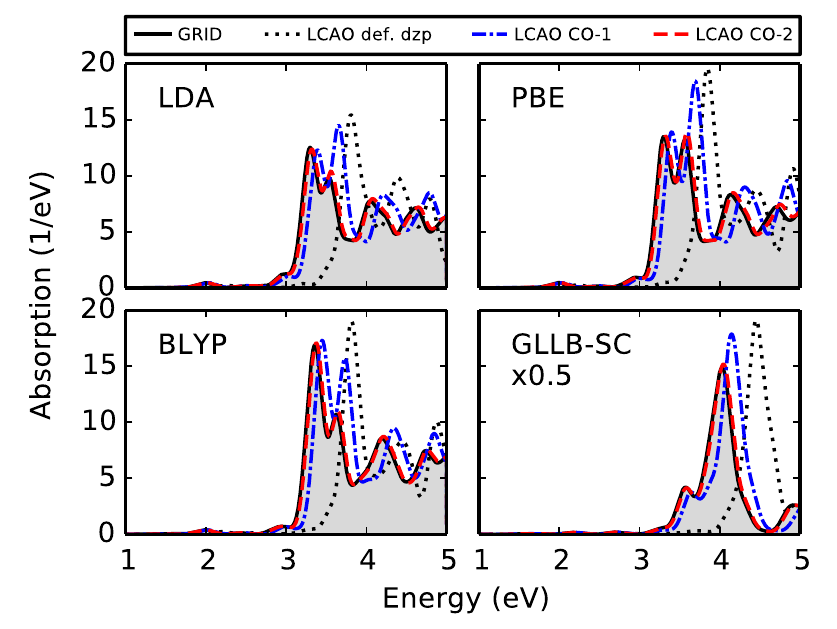}
\caption{Transferability of the PBE-optimized basis sets to different
  xc functionals. Results for \ce{Ag55} are shown. The GLLB-SC
  spectrum has been multiplied by a factor of $0.5$. The
  gray shading is applied to the grid-reference spectra.}
\label{fig xc}
\end{figure}

Different xc functionals yield different shapes of the KS-orbitals. We
studied how the generated NGTOs transfer across various xc
functionals. The basis sets were constructed by augmenting the NAO-sz
basis set corresponding to the chosen functional with the
PBE-optimized NGTOs without any further modification. We used the
\ce{Ag55} cluster as the test system. The results for 
the local density approximation (LDA) \cite{Dirac1930,Bloch1929,Perdew1992}, the Becke-Lee-Yang-Parr (BLYP) functional \cite{Becke1988,Lee1988,Miehlich1989}, and the solid-state modification of the Gritsenko-van Leeuwen-van Lenthe-Baerends potential (GLLB-SC) \cite{Gritsenko1995,Kuisma2010}
are shown in Fig.~\ref{fig xc}, where also the PBE results of Fig.~\ref{fig transf} are
repeated for reference.  While the LDA, PBE, and BLYP functionals
predict similar spectra, the GLLB-SC spectrum has a distinct
shape and stronger intensity.
 This is due to the d-electron screening in coinage metals that is captured
correctly by the GLLB-SC functional. \cite{Kuisma2014,Yan2011First,Yan2012}
The default dzp basis sets, shown for comparison in Fig.~\ref{fig xc}, reproduce 
the strong intensity difference between the GLLB-SC and the other 
functionals but are inadequate to describe the detailed structure of the
spectra, failing to agree with the grid references. In 
contrast, the CO-1 basis sets mostly capture the detailed
differences between the xc functionals,
and the accuracy of the CO-1 basis sets is similar with
all the studied functionals.
The CO-2 results are again in excellent agreement with the grid references in all cases.
Altogether, the results illustrate 
notable transferability of the PBE-optimized CO basis sets to diverse 
xc functionals.

\subsection{Computational performance}

A major advantage of LCAO calculations is that their computational
cost is smaller than that of, \eg, real-space grid
calculations. However, the advantage decreases if large basis sets
must be used. To understand this important issue, we discuss the
effect of basis set size on the computational cost. Fig.~\ref{fig
  time} presents the time-propagation run-times of the generated basis sets
as calculated for the \ce{Ag55} and
\ce{Ag147} clusters with 48 and 96 processors
(cores). \cite{NoteCluster} Depending on the case,
enlarging the basis set from CO-1 to CO-2 (\ie, from $N=17$ to $N=37$)
increases the computational cost by a factor of 2 to 5.

\begin{figure}[b]
\centering \includegraphics{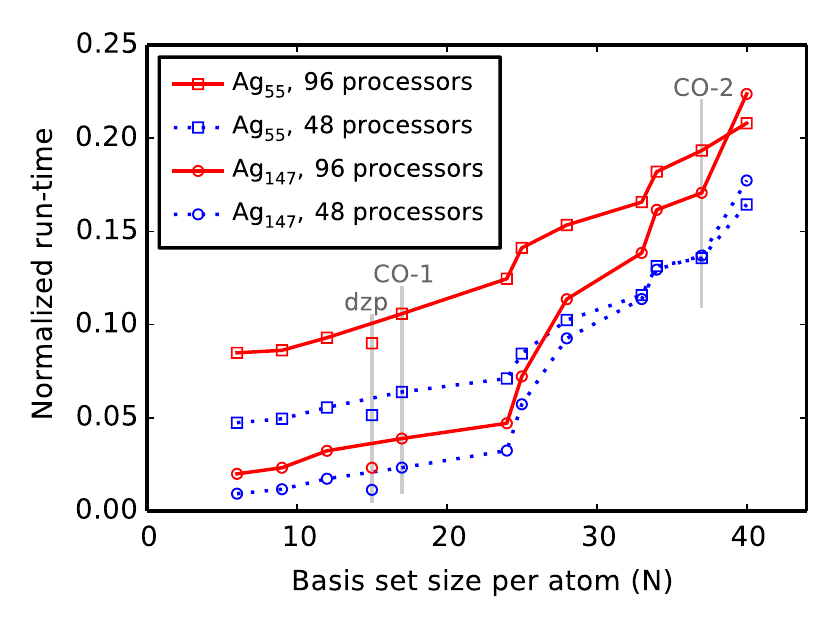}
\caption{Time-propagation run-time of the generated basis set
  series, shown for the \ce{Ag55} and \ce{Ag147} clusters calculated
  with 48 and 96 processors (cores). The run-time of the default 
  dzp basis set is shown for comparison.
  The run-times have been normalized to 
  the run-times of the corresponding grid references 
  calculated with the indicated number of processors.}
\label{fig time}
\end{figure}

The CO-2 calculations are still 5 to 8 times faster than the
grid-reference calculations, while the differences between the results
are minimal as observed in the previous sections. The calculations
with the decently-accurate CO-1 basis set are 10 to 40 times faster
than the corresponding grid calculations. \cite{NoteParall} In the
\ce{Ag147} cluster the speed-ups are in most cases larger than in the smaller
\ce{Ag55} system. This is because the studied systems are relatively
small for the LCAO mode but relatively large for the grid mode, \ie,
when doubling the number of processors from 48 to 96, the grid mode has
excellent scaling, whereas for the LCAO mode the benefit from the
larger number of processors is minor, especially for the small
\ce{Ag55}. Thus, the speed-up factors are expected to be even higher
when the system size is further increased and the LCAO mode is able to
fully take advantage of all the available processors. \cite{Kuisma2014}

In addition to the number of basis functions, the computational
performance is greatly affected also by the spatial extent of the basis functions.
The default dzp basis set does not include diffuse functions that are
important for describing the plasmonic response, leading to smaller run-time than
 the CO basis sets of similar size. 
Within the generated basis set series, the effect of basis function
extent is seen as a staircase-like behavior in 
Fig.~\ref{fig time}. For example, the addition of the f-type NGTO
(\ie, 7 additional basis functions per atom) increases the
computational time only slightly, because the added functions are
short-ranged. On the other hand, the addition of the diffuse s- or
p-type NGTO (\ie, 1 or 3 basis functions per atom, respectively)
affects the computational time more clearly due to the functions'
overlap with functions on nearby atoms. This effect is pronounced in
the larger 147-atom cluster. More aggressive truncation of the NGTOs
might yield better computational performance, but to ensure minimal
deterioration in the accuracy, it may require re-optimization of the
basis sets with the truncation explicitly taken into account in the CO
routine.

\section{Conclusions}
\label{sec conclusions}

In this work, we have addressed the issue of basis set completeness in
time-dependent density functional theory calculations. We have
presented the extension of the completeness-optimization paradigm to
the generation of hybrid NAO+NGTO basis sets, and used it to
parametrize high-accuracy basis sets for nanoplasmonics
calculations. We have demonstrated the performance of the scheme for
the coinage metals Cu, Ag, and Au, which are experimentally
interesting but computationally challenging due to their semi-core
d-electrons that need to be modeled in simulations. We have shown that
the generated basis sets are transferable to simulations of various
metal nanoparticles and nanoalloys as well as to diverse xc
functionals.

The results presented in this work are already promising, but further
improvements of the scheme are still possible. For instance, the error
metric used in the present work may not be optimal. The metric does
not discriminate between different excitations, looking only at the
aggregate intensity. This may result in spuriously small error values
due to interference of different excitations. The use of an error
metric that examines the convergence of the excitations one by one
might yield even better basis set series.

Another approach deserving further development would be to revise the
reference systems against which the basis sets are optimized. In the
present work, accurate and systematically improving basis sets were
obtained by optimizing the basis sets for homoatomic metal atom dimers. The
generated sets were demonstrated to be transferable to larger as well
as heterogeneous systems. However, it might be interesting to optimize
basis sets for extended systems, instead. In a dimer, both atoms are
``on the surface'', which results in the generation of diffuse
functions to model the exponentially decaying density tails. In the
solid state there is no exponential decay, and diffuse functions are
often unnecessary. Nevertheless, our results indicate that the
dynamical response is sufficiently captured already by the dimer for
plasmonics calculations in larger nanoparticles.

The main advantage of the LCAO approach is that satisfactory results
can be obtained much faster than with, \eg, grid-type approaches. The
main problem of the LCAO approach with NAOs (compared to GTO basis
sets of quantum chemistry) has been the scarcity of systematically
better basis sets. This issue has been addressed in the present work.

Although used here for nanoplasmonics, the completeness-optimization
approach is completely general, being applicable to any property at
any level of theory, also beyond DFT (see, \eg,
Refs.~\onlinecite{Lehtola2013Contraction,Lehtola2015}). For this
reason we expect this approach to be widely useful in materials
modeling by electronic structure methods, allowing for large-scale
simulations with better control on their accuracy.

\section{Acknowledgments}

We thank Ask~H.~Larsen for fruitful discussions. T.P.R.\ and
S.L.\ acknowledge financial support from the Vilho, Yrj\"{o} and Kalle
V\"{a}is\"{a}l\"{a} Foundation. S.L.\ also acknowledges the Magnus
Ehrnrooth Foundation for financial support. We thank the Academy of
Finland for support through its Centres of Excellence Programme
(2012--2017) under Project No.\ 251748 and through its FiDiPro
Programme under Project No.\ 263294. We acknowledge the computational
resources provided by the Aalto Science-IT project and CSC -- IT Center
for Science Ltd.\ (Espoo, Finland).

\bibliography{article}

\begin{thebibliography}{97}
\bibitem [{Not({\natexlab{a}})}]{NoteNGTO}%
  \BibitemOpen
  \bibinfo {note} {This NGTO construction scheme is readily available in
  GPAW}\BibitemShut {NoStop}%
\bibitem [{Not({\natexlab{b}})}]{NoteRestr}%
  \BibitemOpen
  \bibinfo {note} {These conditions are equivalent to restricting Gaussian
  exponents to following limits: s-shell: 0.0090--0.93, p-shell: 0.014--1.4,
  d-shell: 0.019--1.9, f-shell: 0.024--2.4}\BibitemShut {NoStop}%
\bibitem [{Not({\natexlab{e}})}]{NoteParall}%
  \BibitemOpen
  \bibinfo {note} {The exact speed-up factors are likely to be affected by the
  choice between the different parallelization schemes}\BibitemShut {NoStop}%
\end{thebibliography}

\end{document}